\documentclass[11pt]{article}
\usepackage{DGfest,epsfig}
\usepackage{times}

\def\Journal#1#2#3#4{{#1} {\bf #2}, #3 (#4)}

\def\NPB{{\em Nucl. Phys.} B}
\def\PLB{{\em Phys. Lett.}  B}
\def\PRL{\em Phys. Rev. Lett.}
\def\PRD{{\em Phys. Rev.} D}
\def\ZPC{{\em Z. Phys.} C}

\def\be{\begin{equation}}
\def\ee{\end{equation}}
\def\bea{\begin{eqnarray}}
\def\eea{\end{eqnarray}}

\newcommand{\beq}{\begin{eqnarray}}
\newcommand{\eeq}{\end{eqnarray}}

\begin{document}
\rightline{DFUP-TH/2006-3, GEF-TH-2006-03}
\baselineskip 11.5pt

\title{TOPOLOGY ON THE LATTICE}

\author{B. All\'es}

\address{INFN, Sezione di Pisa, Largo Pontecorvo 3,
        I--56127 Pisa, Italy}

\author{M. D'Elia}

\address{Dipartimento di Fisica, Universit\`a di Genova, via Dodecaneso 33, I--16146 Genova, Italy}

\maketitle\abstracts{We review the method developed in Pisa 
to determine the topological susceptibility in lattice QCD and present
a collection of new and old results obtained by the method.}

\section{Introduction}

This article is dedicated to the study of topological
properties of QCD on the lattice.
We will describe the method developed in Pisa to measure the topological 
susceptibility on the lattice which is usually known 
as {\em field--theoretical method} and we will review
results obtained by the method together with some recent
applications.

Topology is a difficult subject to be studied on the lattice because
from a strictly mathematical point of view it cannot be defined 
on a discrete space. Its meaning is recovered
in the continuum limit, where the physical
scale gets well separated from the ultraviolet (UV) one.
The topological susceptibility enters in many contexts of QCD
phenomenology~\cite{wit-79,ven-79,shuryak,meggio,ruivo}.
It can be treated as usual expectation values
of field operators in Quantum Field Theory. 
In this approach one does not try to assign
a definite topological sector to each gauge field configuration.
Instead appropriate renormalization constants are used to
relate the lattice averages to the corresponding continuum ones. 
The method developed in Pisa consists in defining the
renormalizations and in giving a prescription to compute  
them on the lattice, as one would do for any other
field--theoretical expectation value. We will review this method 
in Section 2.

In Section 3 we collect some results obtained in the past
by using the field--theoretical method in QCD at zero and finite
temperature, with and without dynamical quarks.
In Section 4  we present new results concerning the application of the
method at finite temperature on anisotropic lattices and at finite
density. A feature of these last results is that several 
determinations of the topological susceptibility are obtained with a fixed
UV cutoff and this fact greatly reduces the number of renormalization
constants required in the computation
and makes the method even more practical and beautiful.

We have written this article to celebrate the 70th birthday of 
Adriano Di Giacomo: the study of topology on the lattice is surely one 
of the leading interests in his research activity, 
an interest which both of the authors have been lucky enough to share with him.

\section{The topological susceptibility}

The topological susceptibility $\chi$ is defined as the zero momentum 
two--point function 
of the topological charge density operator $q(x)$
\beq
\chi &\equiv& \frac{\langle Q^2 \rangle}{V} = 
\int \hbox{d}^4x \; \langle q(x) q(0)  \rangle \; , \;\;\;\;\; Q =
\int  \hbox{d}^4x \; q(x) \; , \\
\label{eq:chi}
q(x) &=& {g^2\over 64\pi^2} \epsilon_{\mu\nu\rho\sigma}
F_{\mu\nu}^a(x)F_{\rho\sigma}^a(x) 
 \; \label{defqx} \; .
\eeq
It contains information about the dependence of the QCD free energy
on the topological $\theta$ parameter around $\theta = 0$,
\beq
\chi = -\frac{1}{V} \frac{d^2}{d \theta^2} \ln Z (\theta) |_{\theta=0}
\; , \;\;\;\;\; Z(\theta) = \int [dA] \; e^{ - \int d^4 x {\cal L_{QCD}}(x)} \;  
 e^{ i \theta\, Q} 
\label{thetaexp}
\eeq
and it enters in various aspects of QCD phenomenology by regulating
the realization of the $U(1)_A$ axial symmetry. For instance, its value in the
pure gauge theory is directly related, at the leading order in
$1/N_c$ ($N_c$ is the number of colours), to the mass of the $\eta'$
meson by the Witten--Veneziano
mechanism~\cite{wit-79,ven-79}, which predicts 
$\chi \approx (180 \, {\rm MeV})^4$.

Topological properties of QCD are of non--perturbative nature: 
lattice simulations are therefore the natural
tool to investigate them.
On the lattice a discretized gauge invariant 
topological charge density operator $q_L(x)$ can be defined, and a related
topological charge $Q_L = \sum_x q_L(x)$, with the only requirement about 
the formal continuum limit $ \lim_{a\to 0} q_L(x) =  a^4 q(x)$
where $a$ is the lattice spacing. A commonly used definition is~\cite{div-81}
\beq
q_L(x) = {{-1} \over {2^9 \pi^2}} 
\sum_{\mu\nu\rho\sigma = \pm 1}^{\pm 4} 
{\widetilde{\epsilon}}_{\mu\nu\rho\sigma} \hbox{Tr} \left( 
\Pi_{\mu\nu}(x) \Pi_{\rho\sigma}(x) \right) \; ,
\label{eq:qlattice}
\eeq
where $\Pi_{\mu\nu}(x)$ is the plaquette operator in the 
$\mu\nu$ plane, ${\widetilde{\epsilon}}_{\mu\nu\rho\sigma}$ is the
Levi--Civita tensor for positive directions and is otherwise
defined by the rule ${\widetilde{\epsilon}}_{\mu\nu\rho\sigma} =
- {\widetilde{\epsilon}}_{(-\mu)\nu\rho\sigma}$.

A proper renormalization must be performed when going towards 
the continuum limit.
In spite of the  formal limit, the discretized topological charge
density renormalizes  multiplicatively~\cite{cdp-88,alvic} (in presence
of dynamical fermions also additive renormalizations can be present,
see next Section)
\beq
 q_L(x) = Z(\beta) a^4(\beta) q(x) + O(a^6) \; , 
\eeq
with a renormalization constant $Z(\beta)$ which is a finite 
function of the lattice bare coupling $\beta = 2 N_c / g_0^2$, approaching
1 as $\beta \to \infty$. 

When defining the topological susceptibility,
further renormalizations  can appear.
Indeed, already the continuum definition of Eq.~(\ref{eq:chi})
involves the product
of two operators $q(x)$ at the same point. Part of this contact term is 
necessary to make $\chi$ a positive quantity 
as required by phenomenology
because $\langle q(x) q(0) \rangle$ is negative for $x \neq 0$
by reflection positivity~\cite{rflps-1,rflps-2,rflps-3,rflps-4,rflps-5,rflps-6}.
Such a term is divergent, so that an appropriate 
prescription must be assigned to define it. This is easily done by
making reference to Eq.~(\ref{thetaexp}), and it corresponds to fixing 
$\chi = 0$ in the sector of zero topological charge.

The lattice definition of the topological susceptibility
\beq
\chi_L = \sum_x \langle  q_L(x) q_L(0) \rangle =  
\frac{\langle Q_L^2 \rangle}{V}
\label{lat_chi}
\eeq
does not generally meet the continuum prescription, leading to the
appearance of an additive renormalization $M$
\beq
\chi_L = Z(\beta)^2 a^4(\beta) \chi + M(\beta) \; .
\label{renchi}
\eeq
The quantity $M(\beta)$ contains the mixing
with all local scalar operators appearing
in the operator product expansion (OPE) of 
$q_L(x) q_L(0)$ as $x \sim 0$ in Eq. (\ref{lat_chi}).

The first lattice determinations~\cite{div-81} of $\chi$ took account
of the mixing with the identity operator
(which gives the main contribution to $M$), but missed
the multiplicative renormalization, so that $Z^2 \chi$ was
measured, obtaining a value quite smaller than predicted by the
Witten--Veneziano mechanism. Based on that,
the idea was put forward that the field--theoretical
discretization of the topological charge
might not be correct and the geometric method~\cite{cite8,cite19},
the cooling method~\cite{cite20,cite21} and Atiyah--Singer based 
methods~\cite{cite22} were developed.
The field--theoretical method was then corrected by
introducing $Z$ and a correct subtraction $M$~\cite{cdp-88,cdpv-90}. 
The development  of a non--perturbative technique, known as the
{\em heating method}, for the 
numerical determination of these constants~\cite{dv-92,prd2284}
finally brought
about a reliable determination of $\chi$, free from the uncontrolled
approximations involved in perturbation theory.

The idea behind the heating method is that the UV fluctuations in
$q_L(x)$, which are responsible for renormalizations,
are effectively decoupled from the background topological
signal so that, starting from a classical configuration of fixed
topological content, it is possible, by applying
a few updating (heating) steps at the 
corresponding value of $\beta$,
to thermalize the UV fluctuations without altering the
background topological content. This is surely true for 
high enough $\beta$, i.e. approaching the continuum limit; 
in practice it turns out to be already true for the values of $\beta$  
usually chosen in Monte Carlo simulations of gauge theories,
being also favoured by the fact that topological modes have very 
large autocorrelation times, as compared to other non--topological observables
(this autocorrelation time is particularly long
in the case of full QCD~\cite{vicari,lippert}).

One can thus create samples 
of configurations with a fixed topological content $Q$ where 
the UV fluctuations are thermalized.
Measurements of topological quantities on those samples can yield
information about the renormalizations. For instance
\beq
\langle Q_L \rangle_Q = Z(\beta) \;  Q \; ,
\label{heat-z}
\eeq
from which the value of $Z(\beta)$ can be inferred, while the expectation
value of $Q_L^2 $ gives 
\beq
\langle Q_L^2 \rangle_Q = Z(\beta)^2 \; Q^2 + V \; M(\beta) \; ,
\label{heat-0}
\eeq
where by $V$ we intend now the four dimensional volume in lattice units
and $\langle \cdot \rangle_Q$ stands for the average within the given
topological sector.

To check that UV fluctuations have been thermalized, one looks
for plateaux in quantities like $\langle Q_L \rangle_Q$ or 
$\langle Q_L^2 \rangle_Q$ as a function of the heating steps performed:
only configurations obtained after the plateau has been reached are 
included in the sample. Special care has to be paid to verify
that during the heating procedure the background
topological charge is left unchanged. This is usually done
by performing a few cooling
steps on a copy of the heated configuration
and configurations where the background topological content 
has changed are discarded from the sample~\cite{fp-94}.
Checks with other methods to measure the background topological
charge (for instance based on operators that satisfy the
Ginsparg--Wilson condition~\cite{gwGW,neubergerGW,lusGW})
lead to perfectly consistent results~\cite{lavoroGW}.

A sample with $Q  = 1 $ can be used to measure $Z$
and a sample with $Q = 0$ (typically thermalized around
the zero field configuration) can be used to determine $M$.
Crosschecks can then be performed, using samples obtained starting
from various configurations with the same 
or different values of $Q$, to test the validity of the 
method~\cite{dv-92,mde-03}.
Once the renormalizations have been computed and the 
expectation value $\chi_L$ over the equilibrium ensemble has been
measured, the physical topological susceptibility $\chi$ is extracted
\beq
\chi = \frac{\chi_L - M(\beta)}{a^4(\beta) Z(\beta)^2}\; .
\label{subchi}
\eeq
An analogous analysis and a similar method to compute the
renormalization
constants can be developed also for higher moments of the 
topological charge distribution~\cite{mde-03}.

If renormalizations are large, i.e. if $Z~\ll~1$
and if $M$ brings a good fraction of the whole signal $\chi_L$, 
the determination of $\chi$ via Eq.~(\ref{subchi})
can be affected by large statistical errors. 
A considerable improvement of the method is thus obtained by using
operators for which the renormalization effects are reduced. 
This is the idea behind the definition of smeared
operators~\cite{cdpv-96},
$q_L^{(i)}(x)$, which are constructed as in Eq.~(\ref{eq:qlattice})
but using, instead of the original links, the 
$i$--times smeared links $U_\mu^{(i)}(x)$ defined as
\beq
U^{(0)}_{\mu}(x) &=& U_{\mu}(x) \; , \nonumber \\
{\overline U}^{(i)}_{\mu}(x) &=& (1-c) U^{(i-1)}_{\mu}(x) +
{c \over 6} 
\sum_{{\scriptstyle \alpha = \pm 1} \atop { \scriptstyle 
|\alpha| \not= \mu}}^{\pm 4}
U^{(i-1)}_{\alpha}(x) U^{(i-1)}_{\mu}(x+\hat{\alpha})
U^{(i-1)}_{\alpha}(x+\hat{\mu})^{\dag}, \nonumber \\
U^{(i)}_{\mu}(x) &=& {{{\overline U}^{(i)}_{\mu}(x)} \over
{ \left( {1 \over N_c} \hbox{Tr} {\overline U}^{(i)}_{\mu}(x)^{\dag} 
{\overline U}^{(i)}_{\mu}(x) \right)^{1/2} } } \; ,
\eeq
where $c$ is a free parameter which can be tuned to optimize the improvement.

\section{Topology at zero and finite temperature in QCD}

The method described in Section 2 has been used in the past 
in various applications; here we will briefly review the main
results. The use of improved smeared operators has been 
essential to obtain high precision measurements, 
especially in the high temperature phase where the vanishing
signal for $\chi$ can get completely lost in Eq.~(\ref{subchi}) 
if renormalizations are large.
Actually previous attempts
to determine $\chi$ across the transition have failed
because the errors involved in the determination were too large with
respect to the physical signal.
The second smearing level has revealed
to be enough both in $SU(2)$~\cite{add-97-2} (with $c = 0.85$) and 
$SU(3)$~\cite{add-97} (with $c = 0.9$) gauge theories. 

\begin{figure}
\begin{center}
\includegraphics[height=5.7cm,width=9.0cm]{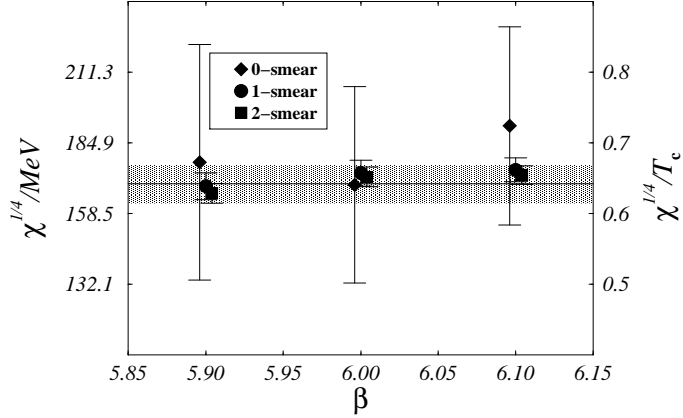}
\end{center}
\caption{Topological susceptibility in $SU(3)$ pure gauge theory calculated
by using 0--, 1-- and 2--smeared lattice topological charge operators (see
text). The central band indicates the average result and its error bar.
Left and right scales are shown in units of MeV and $T_c$.
\label{fig:t0}}
\end{figure}

In Fig.~\ref{fig:t0} we display $\chi$ at zero 
temperature in $SU(3)$ pure gauge theory~\cite{add-97} as
obtained from the measurements on 0--, 1-- and 2--smeared operators at
three different values of the inverse gauge coupling 
$\beta$. The agreement among different operators (universality) 
and the good scaling to the continuum are apparent. From 
the 2--smeared operator the result
$\chi=(170(7)\hbox{ MeV})^4$ is obtained if the value 
$\sqrt{\sigma}=420$~MeV is
considered together with the ratio~\cite{boydengels}
$T_c/\sqrt{\sigma} = 0.629(3)$.
If instead the ratio $T_c/\sqrt{\sigma} = 0.646(7)$
is taken~\cite{luciniteper} then the result is $\chi=(174(7)\hbox{ MeV})^4$.

\begin{figure}
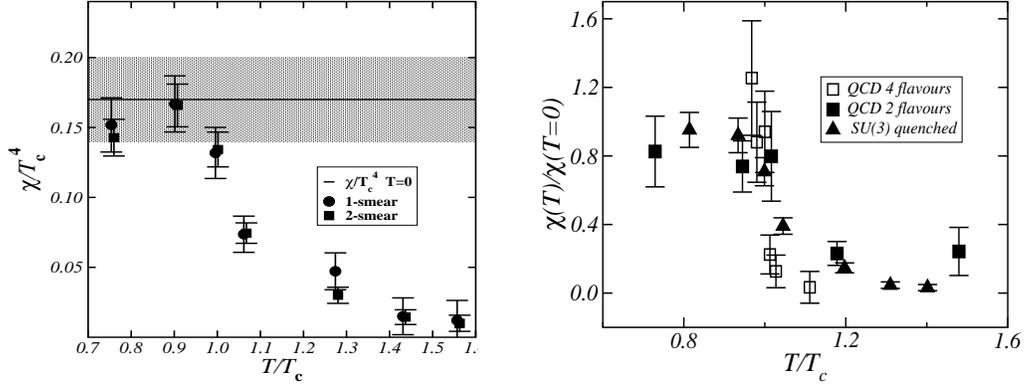

\hskip 0cm
\includegraphics[height=5.1cm,width=6.4cm]{tf.eps}\hspace{0.6cm}
\includegraphics[height=5.1cm,width=6.4cm]{wholeTpureNf4Nf2.eps}
\vspace{0.2cm}
\caption{Topological susceptibility in $SU(3)$ pure gauge theory
(left) calculated by using 1-- and 2--smeared lattice topological
  charge operators (see
text) at finite temperature across the transition at $T_c$.
The central band indicates the average result at zero temperature as
obtained in Fig.~\ref{fig:t0}. A comparison is also made (right) 
with similar results obtained in full QCD with $N_f = 2$ and
$N_f = 4$.
\label{fig:tf}}
\end{figure}

The behaviour of $\chi$ across the finite temperature transition is an
important ingredient to understand the fate of the singlet axial symmetry
at deconfinement and/or chiral $SU(N_f)$ restoration\cite{meggio,ruivo}.
The behaviour of $\chi$
across the transition has been obtained~\cite{add-97} and it is shown
in Fig.~\ref{fig:tf}.
 The susceptibility stays constant until the
deconfinement temperature $T_c$ and then it undergoes an abrupt drop as
shown in the Figure.
Similar results are obtained in $SU(2)$ pure gauge theory~\cite{add-97-2}
and for the unquenched theory~\cite{plb483} with 2 and 4 staggered
quarks. In Fig.~\ref{fig:tf} the behaviours of the quenched,
$N_f=2$ and $N_f=4$ theories with gauge group $SU(3)$ are put
together for comparison.
In the unquenched calculation one has to take into account that
the topological charge
can also mix with other operators related to the 
anomaly~\cite{espiru}.
Usually we neglect this mixing because it is rather small and
surely smaller than the error bars from the simulation~\cite{plb350}.

Another problem that has been studied by using the field--theoretical
method is the possible spontaneous breaking of parity in Yang--Mills
theory. An old theorem by Vafa and Witten
precluded such a possibility~\cite{vafawitten}.
However the authors implicitly assumed
the breaking to be absent when they imposed that the derivative
of the free energy with respect to $\theta$ is zero at the
minimum of the function $ - \ln Z(\theta)$.
Our goal was 
to obtain a bound from lattice on the
parity breaking effects, in particular to the electric dipole moment
of neutral baryons: 
that was done by looking at a possible volume 
dependence of the topological susceptibility.
After a thorough simulation on rather large volumes (up to $48^4$) and 
huge statistics the following bound on the neutron electric
dipole moment was obtained~\cite{theta} 
(although it is still almost 5 orders of
magnitude less precise than the corresponding
experimental limit): $ d_e 
    <  3.5\; 10^{-21} \; e\cdot{\rm cm}\;.$
It must be stressed
that the field--theoretical
method enabled us to use such a large statistics and lattice sizes,
a goal that would be hard to reproduce with other methods.

\section{Recent applications of the field--theoretical method}

In the applications of the Pisa method that we have reviewed in 
Section 3, a determination of $\chi$ was usually needed at several
values of the inverse gauge coupling $\beta$, involving several
different determinations of the renormalization constants 
$Z(\beta)$ and $M(\beta)$.

Suppose now that we want to study the behaviour of
$\chi(\alpha)$ on some parameter $\alpha$, which does not affect the 
UV behaviour of the theory, while $\beta$ and other bare 
parameters (like the masses of dynamical fermions,
if present) are kept fixed.
This happens in some interesting cases:
we will consider the study of 
the behaviour of $\chi$ across the finite temperature deconfining
transition on anisotropic lattices at fixed $\beta$ and variable 
temporal extension (in this case $\alpha$ is the
temporal extent $N_t$ of the lattice) and the  determination
of $\chi$ across the finite density deconfining transition at fixed
temperature (in this case $\alpha$ is the chemical potential $\mu$).
Renormalization constants are generated by quantum fluctuations 
at the UV scale, therefore they are independent of $\alpha$ and we can write:
\beq
\chi(\alpha) = \frac{\chi_L(\alpha,\beta) - M(\beta)}{a^4(\beta) Z(\beta)^2}\; .
\label{subchialpha}
\eeq
In the right hand side of Eq.~(\ref{subchialpha}) $\chi_L$ is the only
quantity which depends on $\alpha$:
if one is not interested in an overall multiplicative factor, 
which is the case 
when studying the critical behaviour across a phase transition, 
the determination of a single additive renormalization constant
is all that is needed to study the dependence of $\chi$ on $\alpha$.
In those cases the field--theoretical method is much simpler and
less computer--time consuming than other methods.

\subsection{Topology on anisotropic lattices}

In the path integral approach to Quantum Field Theory,
a finite temperature can be introduced 
by fixing a finite temporal extent $\tau$ in the Euclidean
space--time
with (anti)--periodic boundary conditions for (fermionic)
bosonic fields, the temperature being $T = 1/\tau$. On a lattice
this expression becomes $T = 1/(N_t a(\beta))$.
The temperature can thus be varied by changing either the number of 
lattice sites in the temporal direction or the lattice
spacing, hence $\beta$ (we are considering the pure gauge case, otherwise
$a$ would depend on the bare quark masses as well).
Usually the second option is chosen: indeed for $T$ near the QCD
phase transition ($T_c \sim 200$ MeV) and for the lattice spacings
affordable by present computers ($a^{-1} \sim 1-2$ GeV) typical
values of $N_t$ are too small to allow a fine tuning of $T$
around $T_c$ by only varying the temporal length.

The situation changes when using anisotropic
lattices~\cite{klassen}, 
where different bare couplings are used in temporal and spatial
planes, leading to different lattice spacings $a_s$  and 
$a_t$.
One can thus use a very fine temporal spacing 
while leaving the spatial ones coarse, with an affordable 
computer time cost. Anisotropic lattices have been originally introduced
to tackle problems related to heavy quarks, glueballs and high
temperature thermodynamics which are not easily manageable
otherwise.
Here we will exploit the possibility of fine tuning
$T$ around $T_c$ 
by simply changing $N_t$ and leaving $a_s$ and $a_t$ fixed.

We will consider the pure gauge plaquette action for $N_c = 3$,
\beq
	S_G=
	{\beta \over N_c}
	{1 \over \gamma_{\rm }}
	\sum_{x, i<j \le 3}
	\mbox{Re}\mbox{Tr}
	\left( 1 - \Pi_{ij}(x) \right)
	+
	{\beta \over N_c}
	\gamma_{\rm}
	\sum_{x, i \le 3}
	\mbox{Re}\mbox{Tr}
	\left( 1 - \Pi_{i4}(x) \right) \; .
\label{anisotchange}
\eeq
Both $a_s$ and $a_t$, as well as the renormalized anisotropy
$\xi = a_s/a_t$,  are functions of $\beta$ and of the
bare anisotropy $\gamma$. Several determinations of those 
parameters can be found in the literature, we will refer 
to the following values~\cite{klassen,ishii}: $\beta = 6.25$,
$\gamma = 3.2552$, which leads to $\xi = 4$ and $a_s \approx 0.021$
fm. The critical temperature is reached for $N_{t,c} \approx 35$, so that
simulating at a different $N_t$ corresponds to $T/T_c = N_{t,c}/N_t$. 
We have chosen a spatial lattice size $N_s = 24$ which corresponds
to a spatial extent of about 2 fm.

\begin{figure}
\begin{center}
\includegraphics[height=5.7cm,width=8.0cm]{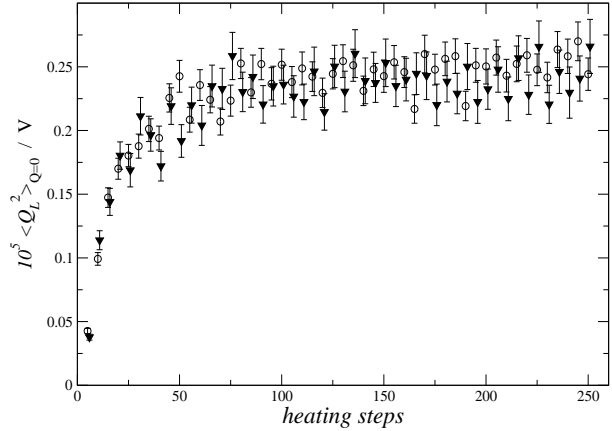}
\end{center}
\caption{Comparison
of the results obtained for $\langle Q_L^2 \rangle_{Q = 0} / V$ as a function 
of the heating steps performed for the operator
at the second smearing level on anisotropic lattices of two different
temporal extensions: $N_t = 20$ (circles) and $N_t = 40$ (triangles)
.
\label{fig:Manisot}}
\end{figure}

\begin{figure}
\begin{center}
\includegraphics[height=5.7cm,width=7.0cm]{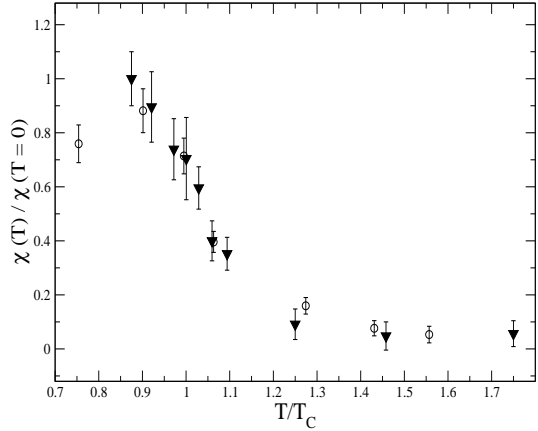}
\end{center}
\caption{Topological susceptibility as a function of $T/T_c$ as 
obtained on anisotropic lattices by using the operator at the second
smearing level (triangles) compared with previous
results~\protect\cite{add-97}
obtained using the same operator on isotropic lattice (circles).
\label{fig:Chianisot}}
\end{figure}

We have determined $\chi(T)/\chi(0)$ using the topological charge
operator at the second
smearing level and compared with previous results obtained on
isotropic lattices. 
As a first step we have checked that the additive renormalization
constant
is indeed independent of the temporal extent $N_t$. We have computed
$M$ for two different temporal extensions, $N_t = 20$ and $N_t = 40$, 
corresponding respectively to $T/T_c \approx 1.75$ and $T/T_c \approx
0.875$. We have obtained $M = (0.249 \pm 0.010) \cdot 10^{-5}$ for 
$N_t = 20$ and $M = (0.242 \pm 0.009) \cdot 10^{-5} $ for
$N_t = 40$, i.e. a very good agreement which is also apparent directly
from Fig.~\ref{fig:Manisot}, where we report 
$\langle Q^2_L \rangle_0 /V$ as a function of the heating step. An
average of the two determinations of $M$ has been used for other
values of $N_t$.

In Fig.~\ref{fig:Chianisot} we report the quantity
$\chi(T)/\chi(0) = (\chi_L(T) - M)/(\chi_L(0) - M)$ obtained
for the 2--smeared operator and compare it with the same quantity
determined on isotropic lattices~\cite{add-97}. In the present case 
$\chi(0)$ is not available and it has been fixed to the value 
obtained at the lowest available temperature, $T = 0.875 T_c$. A quite
good agreement is visible.

\subsection{Topology at finite density}

An analogous situation is met when varying the chemical potential 
$\mu$ at constant temperature, which means at fixed $\beta$,
$N_t$ and quark mass values: 
indeed a finite $\mu$ does not 
affect the dynamics at the UV scale, at least until $\mu$ is
small with respect to the UV cutoff (for very large values 
of $\mu$ Pauli blocking sets up with a consequent quenching of 
fermion dynamics at all scales).
We report in this Section 
results obtained~\cite{su2dens} for the theory with $N_c = 2$,
which is the only case where the sign problem does not make
Monte Carlo simulations at finite $\mu$ unfeasible. A lattice
of spatial size $N_s = 14$ and temporal size $N_t = 6$ has been
used, with dynamical staggered fermions corresponding to 8
continuum degenerate flavours of bare mass $a  m = 0.07$
and an inverse coupling $\beta = 1.5$, corresponding to
a temperature well below $T_c$: a phase transition to deconfined
matter is therefore expected after increasing $\mu$ beyond some 
critical value $\mu_c$. The aim was to investigate
the fate of topological excitations, hence of $\chi$, across
the finite density phase transition.

\begin{figure}
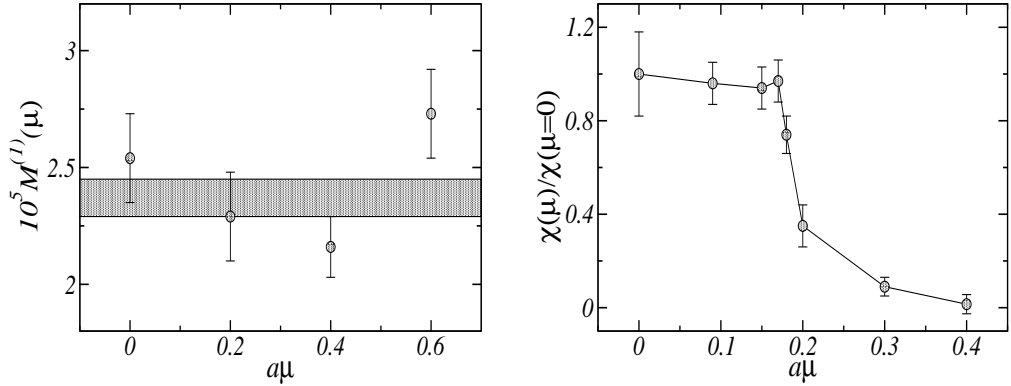

\hskip 0cm
\includegraphics[height=5.1cm,width=6.3cm]{M1mu.eps}\hspace{0.6cm}
\includegraphics[height=5.1cm,width=6.3cm]{topmu_SOLO.eps}
\vspace{0.2cm}
\caption{Additive renormalization (left) and $\chi(\mu)/\chi(0)$ (right)
as a function of $a\mu$ for the theory at finite density 
with 2 colours and 8 continuum dynamical flavours.
\label{fig:Chimu}}
\end{figure}

In the l.h.s. of Fig.~\ref{fig:Chimu} the estimates of $M$
obtained for four different values of $\mu$ are shown.
They look compatible within errors as they should be
if the renormalization procedure is correct and no
density effects are introduced into the subtractions.
Hence also in this case a single renormalization
constant is enough to determine the ratio
$\chi(\mu)/\chi(0) = (\chi_L(\mu) - M)/(\chi_L(0) - M)$ which
is displayed in the r.h.s. of Fig.~\ref{fig:Chimu}. It shows a
clear drop at a critical value $\mu_c$, analogously to what
happens for the finite temperature phase transition.

\section*{Acknowledgments}
The authors are greatly indebted to Adriano Di Giacomo for
his acumen, the precious scientific collaboration and his sense of humanity.
They also acknowledge collaboration with M.P. Lombardo and M. Pepe 
for the results obtained at finite density.

\end{document}